\documentclass[11pt]{article}
\usepackage{natbib,bm,amssymb, amsmath}
\usepackage{graphicx}
\usepackage{color,xcolor,setspace,geometry}
\usepackage{diagbox,comment}
\usepackage{url}
\usepackage{arydshln}
\usepackage{pdflscape}
\usepackage{adjustbox}

\usepackage{rotating}
\usepackage[titletoc,title]{appendix}

\begin{document}

\begin{spacing}{1.35}	

\title{Extending the Dixon and Coles model: an application to women's football data}

\author{Rouven Michels\thanks{corresponding author: r.michels@uni-bielefeld.de} \thanks{ Bielefeld University}, Marius \"Otting\footnotemark[2], Dimitris Karlis\thanks{Athens University of Economics and Business}}
\date{}

\maketitle

\begin{abstract}
The prevalent model by \citet{dixon1997modelling}  
extends the double Poisson model where two independent Poisson distributions model the number of goals scored by each team by 
moving probabilities between the scores 0-0, 0-1, 1-0, and 1-1. 
We show that this is a special case of a multiplicative model known as the Sarmanov family.  
Based on this family, we create more suitable models by moving probabilities between scores and employing 
other discrete distributions.  
We apply the new models to women's football scores,  
which exhibit some characteristics different than that of men's football. 
\end{abstract}

\section{Introduction}
Football is the most popular sport in the world. There is an ever-growing interest in predicting the scores of football matches for various purposes, including betting, improved team organisation and just fun among other reasons. In the academic literature, \citet{maher1982modelling} was the first to systematically investigate the the number of goals scored in men's football. Since his seminal work, many extensions have been proposed, aiming at modelling the number of goals scored by each team. When modelling this number, two main questions arise. First, one has to select a marginal distribution of the number of goals --- the Poisson distribution constitutes a standard choice. Second, the correlation between the number of goals scored by the two competing teams is also essential. Commonly used models
include the double Poisson model of \citet{lee1997modeling}, bivariate Poisson models as in \citet{maher1982modelling, dixon1997modelling, karlis2003analysis, baio2010bayesian, groll2018dependency}, and models based on copulas defined in \citet{mchale2007modelling}, which also allow for marginals others than Poisson, e.g., the negative binomial distribution \citep{mchale2011modelling} or the Weibull count distribution \citep{boshnakov2017bivariate}. 

Among these models, the one proposed by \citet{dixon1997modelling} has found tremendous impact. In particular, \citet{dixon1997modelling} found that scores like 0-0s and 1-1s are more likely to occur with real data than
under an independence assumption. To address this characteristic, they developed a model that shifts probability between the scores 0-0, 0-1, 1-0, and 1-1. As their approach has proven helpful in modelling football scores, it is considered one of the most widely used models. However, in their model formulation, it is not possible to shift probabilities of scores other than 0-0, 0-1, 1-0, and 1-1 and to use marginals other than Poisson. If only these four mentioned scores occur more (or less) likely than under independence, there is no need to shift the probabilities of the other scores. In fact, in men's football, the empirical proportions of other common scores, such as 2-0 and 3-0, are usually very close to what would be expected under independence. However, 
this is usually not the case for women's football. 

In women's football, scorelines such as 2-0 and 3-0 are much more likely to occur than under independence. However, we cannot modify the corresponding probabilities for such scores in the Dixon and Coles model. To overcome this limitation and to adequately model women's football scores, we extend the Dixon and Coles model in several ways. In particular, we first show that this model is a special case of the Sarmanov family of distributions \citep{sarmanov1966generalized}. Second, exploiting the connection to the Sarmanov family of distributions, we demonstrate how to shift probabilities of scores other than 0-0, 1-0, 0-1, and 1-1. In particular, even an infinite number of probabilities can be modified. Third, we allow for marginal distributions other than Poisson. Fourth, since the correlation implied under the Dixon and Coles model is relatively small and thus unrealistic in some applications, we present model formulations which are applicable to data with a wider range of correlation. 

To demonstrate the feasibility of our approach, we consider data on the number of scored goals in the four most popular women's football leagues in Europe. In particular, we consider the English FA Women's Super League, the German Frauen-Bundesliga, the French Division 1 Feminine, and the Spanish Primera Iberdrola for the seasons 2011/12--2018/19 and 2021/22. Interest in women's football has increased in recent years as nowadays, demand for several women's football matches is as high as for men's football. In March 2022, more than 90,000 spectators witnessed the Champions League quarter-final between FC Barcelona and Real Madrid in the Camp Nou. More than half a million people attended the UEFA Women's Euro 2022. Despite this increased interest in women's football, it has been analysed very briefly to date, with most studies comparing men's and women's football (see, e.g.\ \citealp{martinez2014women, pollard2014comparison, pedersen2019scaling, pappalardo2021explaining, garnica2021understanding}). To our knowledge there is limited research on modelling women’s football scorelines in contrast to the men's game.

The rest of the paper is structured as follows. Section 2 introduces the model proposed by \citet{dixon1997modelling} and presents several extensions. Section 3 covers the application of the models presented in Section 2 to the women's football data, thereby comparing their fit and predictive performance. Section 4 concludes. 

\section{Extending the Dixon and Coles model}
In this section, we show that the model developed by \citet{dixon1997modelling} is a special case of a vast family of probability distributions for modelling bivariate count data, namely the Sarmanov family of distributions (\citealt{sarmanov1966generalized}). This relationship implies that we can extend the Dixon and Coles model in specific directions, including other discrete distributions than Poisson or other dependence structures by altering the function that introduces correlation.

\subsection{The Dixon and Coles model}
\citet{dixon1997modelling} defined a bivariate model with Poisson marginal distributions. The corresponding joint probability mass function (pmf) is given by
\[
P(X_1 = x_1, X_2 = x_2) = \tau_{\lambda_1, \lambda_2}(x_1,x_2)\frac{{\lambda^{x_1}_1} \exp ( -\lambda_1 )}{x_1!}\frac{{\lambda^{x_2}_2} \exp (-\lambda_2 )}{x_2!},
\]
with
\[
\tau_{\lambda_1, \lambda_2}(x_1,x_2) = \left\{ \begin{array}{cc}
1- \lambda_1 \lambda_2 \tilde{\omega}, \ & \mbox{if}~ x_1=x_2=0, \\
1 + \lambda_1 \tilde{\omega}  \ & \mbox{if}~ x_1=0,\ x_2=1, \\
1 + \lambda_2 \tilde{\omega}  \ & \mbox{if}~ x_1=1,\ x_2=0, \\
1 - \tilde{\omega}  \ & \mbox{if}~ x_1=x_2=1, \\
1 & \mbox{otherwise}\,,
\end{array}
\right.
\]
where $\lambda_1$ and $\lambda_2$ are the means of the two Poisson marginal distributions and $\tau_{\lambda_1, \lambda_2}(\cdot,\cdot)$ measures the correlation between the scores. For $x_1 > 1$ and $x_2 >1$, the probabilities remain unchanged from the product of the marginal distributions as probabilities are shifted only between the four pairs $(0,0)$, $(1,0)$, $(0,1)$ and $(1,1)$. The magnitude of shifting depends on the dependence parameter $\tilde{\omega}$, which has to satisfy the following inequality: 
\[  \max{(-1/\lambda_1, -1/\lambda_2)}\leq \tilde{\omega} \leq \min{(1/\lambda_1\lambda_2,1)}. \]
Thus, $\tilde{\omega}$ can take positive or negative values but it is of limited range for reasonable means $\lambda_1$ and $\lambda_2$. The case $\tilde{\omega} =0$ corresponds to scores being independent. 

\begin{figure}
    \centering
    \includegraphics[scale=0.7]{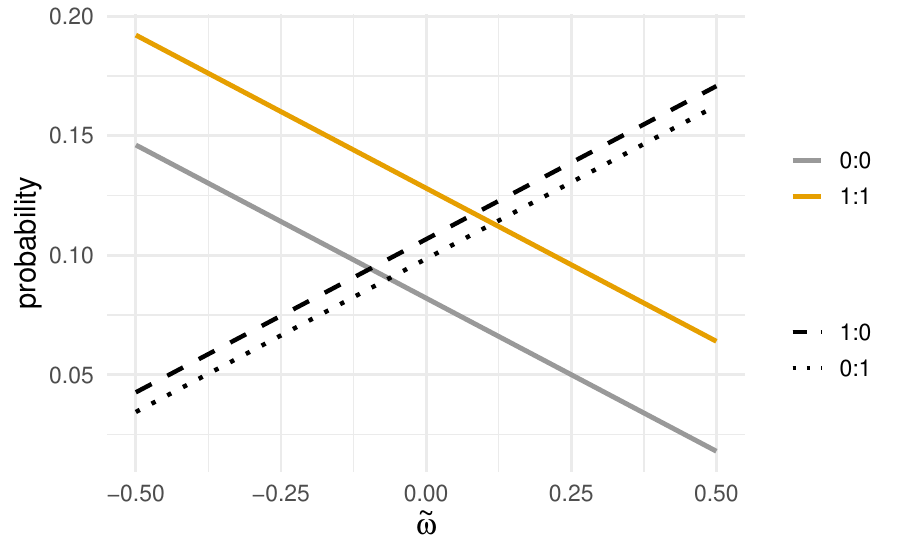}
    \caption{Effect of the dependence parameter $\tilde{\omega}$ in the Dixon and Coles model on the probabilities for the pairs $(0,0)$, $(1,0)$, $(0,1)$ and $(1,1)$. The means,  of the Poisson marginals are fixed at $\lambda_1 = 1.3$ and $\lambda_2 = 1.2$, respectively.}
    \label{fig:effect_rho}
\end{figure}

How the probabilities for the pairs $(0,0)$, $(1,0)$, $(0,1)$ and $(1,1)$ are affected by $\tilde{\omega}$ is shown in Figure~\ref{fig:effect_rho}. In particular,
for increasing values of $\tilde{\omega}$, probabilities under the Dixon and Coles model are shifted from $(0,0)$ and $(1,1)$ to the pairs $(0,1)$ and $(1,0)$ in a proportional manner.

\subsection{Sarmanov Family}
The Sarmanov family was introduced by \citet{sarmanov1966generalized}, while \citet{ting1996properties} studied some general methods for the construction of different families considering different types of marginal distributions. Throughout this contribution, we focus on the case of discrete distributions. 

For $i =1,2$, assuming $P_{i}(x_{i})$ are two probability mass functions (pmf)
and $q_{i}(x_{i})$ are two bounded non-constant functions such that
    \begin{equation}\label{cond}
        \sum\limits_{x_{i}=-\infty}^{\infty} q_{i}(x_{i}) P_i(x_{i})=0,
    \end{equation}
then a joint pmf can be defined by
    \begin{equation}
    P(X_1 = x_{1},X_2 = x_{2})=P_{1}(x_{1})P_{2}(x_{2})[1+ \omega q_{1}(x_{1})q_{2}(x_{2})],
    \label{eq4}
    \end{equation}
where $\omega q_{1}(x_{1})q_{2}(x_{2})$ specifies the dependence of $X_{1}$ and $X_{2}$ and $\omega \in \mathbb{R}$ satisfies, for all $x_1$ and $x_2$, the condition
        $[1+ \omega q_{1}(x_{1})q_{2}(x_{2})] \geq 0.$
For $\omega=0$, the variables $X_{1}$ and $X_{2}$ are independent.

Following \citet{ting1996properties}, the correlation between $X_1$ and $X_2$ is then given by $$\rho = \frac{\omega u_1 u_2}{\sigma_1 \sigma_2}$$
where 
$\sigma_i$ is the standard deviation of the marginal distribution and $u_i = E[X_i q_i(X_i)]$ for $i = 1,2$.

\subsection{Dixon and Coles model as a member of Sarmanov Family}
By selecting suitable functions $q_i(x_i)$ that fulfil equation (\ref{cond}), we can build flexible bivariate distributions based on the Sarmanov family. The model by \citet{dixon1997modelling} introduced in Section 2.1 also belongs to the Sarmanov family. For Poisson marginals as considered by \citet{dixon1997modelling}, we set $\omega = -\tilde{\omega}$ and select the functions $q_1(x_1)$ and $q_2(x_2)$ as
\begin{equation*}
q_{dc}(x_i) = \left\{ \begin{array}{ll}
-\lambda_i & \text{if } x_i=0 \\
1 &  \text{if } x_i=1 \\
0 &  \text{if } x_i=2,3,\ldots
\end{array}   \right.
\end{equation*}
where $\lambda_i$ is the mean of the variable $X_i$, for $i = 1,2$. Then, the condition in equation (\ref{cond}) holds when assuming Poisson marginals. In fact, plugging $q_{dc}(x_i)$ into equation (\ref{eq4}) yields the pmf of the model proposed by \citet{dixon1997modelling}. 

To create more flexible bivariate discrete distributions with Poisson marginals, we may use other $q-$functions such that these functions still fulfil equation (\ref{cond}). To our best knowledge, we are the pioneers in extending and enhancing the Dixon and Coles model within the Sarmanov family. In the following subsection, we provide such extensions.

\subsection{Some new models}
\subsubsection{Poisson Marginals}
Note that throughout this subsection, we will consider $\lambda_i$ as the mean of a Poisson distributed variable $X_i$, $i~=1,2$.
Still assuming Poisson marginal distributions, we can consider another function $\hat{q}$ defined as
\begin{equation*}
\hat{q}(x_i) = \left\{ \begin{array}{ll}
-\lambda_i^2 & \text{if } x_i=0 \\
\lambda_i & \text{if } x_i=1 \\
0 & \text{if } x_i=2,3,\ldots
\end{array}   \right.
\end{equation*}
for $i = 1,2$, 
which also satisfies equation (\ref{cond}). 
This generates another bivariate distribution with Poisson marginals, but probabilities across the four pairs $(0,0)$, $(1,0)$, $(0,1)$ and $(1,1)$ are now shifted differently. In particular, we shift probabilities with a quadratic term. However, we can formulate $\hat{q}$ also with other exponents for $\lambda_i$ --- in Appendix \ref{sec:app_other_exponents}, equation (\ref{newq2}) provides such a generalisation.

A peculiarity of the $q-$functions presented so far is that they shift probabilities only across the four pairs $(0,0)$, $(1,0)$, $(0,1)$ and $(1,1)$. However, we can easily relax this restriction. For example, we can extend the previous model to $x_i = 2$ with the following $q-$function:
\[
\Tilde{q}(x_i) = \left\{ \begin{array}{ll}
-\lambda_i^2 & \text{if } x_i=0 \\
-\lambda_i & \text{if }  x_i=1 \\
4 & \text{if } x_i=2 \\
0 & \text{if } x_i=3,4, \ldots
\end{array}   \right.
\] for $i = 1,2.$

Such a function moves the probabilities of the pairs $(x_1,x_2): x_1=0,1,2;~x_2=0,1,2$ and thus induces correlation among nine pairs. We can extend this further 
up to $x_i=s$, i.e., inducing correlation among $(s+1)^2$ pairs, by considering the following general function
\[
q^{(s)}(x_i) = \left\{ \begin{array}{ll}
-x_i! \lambda^{s-x_i} & \text{if } x_i=0,1,\ldots,s-1 \\
s s! & \text{if } x_i=s \\
0 & \text{if } x_i=s+1,\ldots 
\end{array}   \right.
\]
for $i = 1,2$.

So far, we have always selected the same $q-$function for each marginal distribution. However, we can also consider different functions $q$ for each marginal distribution. For example, considering $q^{(2)}(x_1)$
and $q^{(3)}(x_2)$, 
we can alter the probabilities for a broader range of values, in this example for all pairs $(x_1,x_2): x_1=0,1,2,~x_2=0,1,2,3$.
Note that all the above-mentioned $q$ functions satisfy equation (\ref{cond}) and thus they provide proper bivariate discrete distributions, while they also define the appropriate limits for the $\omega$ parameter.

\begin{figure}
    \centering
    \includegraphics[width = \textwidth]{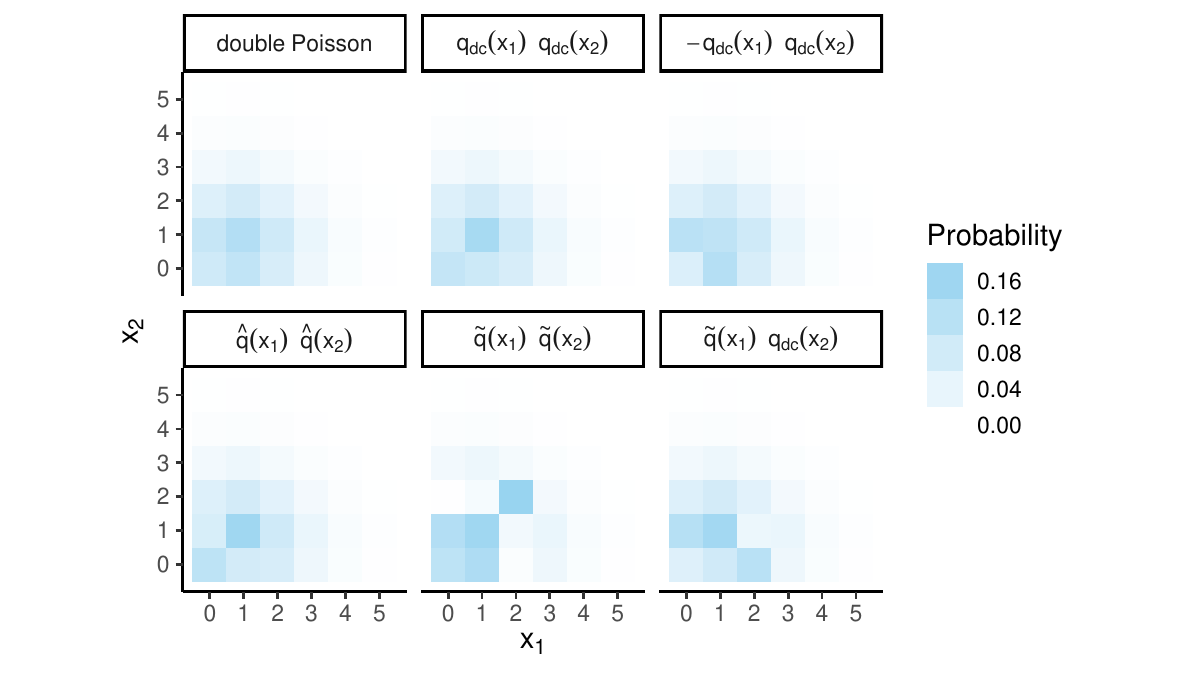}
    \caption{Different (discrete) bivariate distributions with Poisson marginals based on the Sarmanov family. 
    For all panels, we consider Poisson marginals with means 1.3 and 1.2, respectively and a dependence parameter of 0.15.} 
    \label{fig:probabilities}
\end{figure}

The panels in Figure \ref{fig:probabilities} show the probabilities under the different model formulations, i.e.\ the model under independence, the model proposed by \citet{dixon1997modelling} and models using the different $q-$functions developed so far.

\subsubsection{Any other discrete distribution} 
In the previous subsection, we demonstrated how we can extend the Dixon and Coles model for Poisson marginals using other $q-$functions. It is relatively straightforward to proceed with similar functions for the case of any discrete distribution. Consider, for example, a discrete distribution function with probability assigned to $x_i$ given as $P_{x_i}$, $x_i \in \mathbb{N}_0$. 
Let $\mu_i$ denote the expected value of this pmf. Remembering the assumption in equation (\ref{cond}) that needs to be satisfied, we can generalise the form of the Dixon and Coles model by creating some new $q-$functions for general discrete distributions as
\[
q_{1P}(x_i) = \left\{ \begin{array}{ll}
-\frac{P_1}{P_0} & \text{if }x_i=0 \\
1 &  \text{if }x_i=1 \\
0 &  \text{if }x_i=2,3, \ldots
\end{array}   \right. 
~~\mbox{or}~~ 
q_{2P}(x_i) = \left\{ \begin{array}{ll}
\mu_i & \text{if }x_i=0 \\
-\mu_i \frac{P_0}{P_1}  &  \text{if }x_i=1 \\
0 &  \text{if }x_i=2,3, \ldots
\end{array}   \right.
\]

\noindent For a function equivalent to $\hat{q}$ we can define
\[
q_{3P}(x_i) = \left\{ \begin{array}{ll}
-\mu_i \frac{P_1}{P_0} & \text{if }x_i=0 \\
 \mu_i  &  \text{if }x_i=1 \\
0 &  \text{if }x_i=2,3, \ldots
\end{array}   \right.
\]
for $i = 1,2$.
For Poisson marginals, since $\mu_i=\lambda_i$ and $P_1/P_0 =\lambda_i$, we can verify that the resulting functions are identical to those presented in the previous subsection. Up next, we will present a slightly more involved example with negative binomial margins.

\subsubsection*{Example: negative binomial marginals}
For our application of modelling the number of goals in football, the Dixon and Coles model considers only Poisson marginals, which can be very restrictive for football modelling. To this end, we present an extension to negative binomial distributed variables $X_1$ and $X_2$. 
To find a model that shifts probabilities only across the four pairs $(0,0)$, $(1,0)$, $(0,1)$ and $(1,1)$ --- similar to the Dixon and Coles model --- and that meets the constraints imposed by the Sarmanov family in equation (\ref{cond}), we use the $q_{1P}(x_i)$ function presented in the previous subsection 2.4.2. Thus, with $P_1~=~P(X_i~=~1)~=~ \phi_i \left(\frac{\mu_i}{\phi_i + \mu_i}\right) \left(\frac{\phi_i}{\phi_i + \mu_i}\right)^{\phi_i}$ and $P_0~=~P(X_i = 0)~=~\left (\frac{\phi_i}{\phi_i + \mu_i}\right)^{\phi_i}$,
$i=~1,2$, we end up with the following function:
\begin{equation*}
q_{nb}(x_i) = \left\{ \begin{array}{ll}
-\phi_i\ \left(\frac{\mu_i}{\phi_i + \mu_i}\right) & \text{if }x_i=0 \\
1 &  \text{if }x_i=1 \\
0 &  \text{if }x_i=2,3, \ldots
\end{array}   \right.
\end{equation*} 
with $\mu_i$ denoting the mean and $\mu_i + \frac{\mu_i^2}{\phi_i}$ the variance of the negative binomial distribution, for $i=1,2$.
This function moves probabilities for $x=0$ and $x=1$ as in the Dixon and Coles model. 
However, similar to Poisson marginals, we can build further $q-$functions for negative binomial marginals such as
\[
\hat{q}_{nb}(x_i) = \left\{ \begin{array}{ll}
-\mu_i^2 & \text{if }x_i=0 \\
\mu_i \frac{\phi_i + \mu_i}{\phi_i} &  \text{if }x_i=1 \\
0 &  \text{if }x_i=2,3, \ldots
\end{array}   \right.
~~\mbox{or}~~ 
\Tilde{q}_{nb}(x_i) = \left\{ \begin{array}{ll}
-\mu_i^2 & \text{if }x_i=0 \\
-\mu_i \frac{\phi_i + \mu_i}{\phi_i} &  \text{if }x_i=1 \\
4 \mu_i \frac{\phi_i}{\phi_i + \mu_i} &  \text{if }x_i=2 \\
0 &  \text{if }x_i=3,4, \ldots
\end{array}   \right.
\]
for $i = 1,2.$

Finally, note that such derivations are also valid when considering marginal distributions from two different families. In particular, we can define discrete bivariate distributions by selecting suitable $q-$functions according to the marginal distributions. 
For example, we can consider equation (\ref{eq4}), with $P_1(x_1)$ and $P_2(x_2)$ as the pmfs of the Poisson and negative binomial distribution, respectively, and $q_1(x_1)$ and $q_2(x_2)$ given by $q_{dc}(x_1)$ and $q_{nb}(x_2)$, respectively.
Such constructions can provide more powerful bivariate discrete distributions. 

\subsection{Shifting probabilities across the entire support}
Up to this point, we only investigated bivariate distributions shifting probabilities for a pre-defined set of values.
However, some applications require to relax this assumption such that we shift probabilities more flexibly between all values of the support of the marginal distributions. 
To set up a model based on the Sarmanov family that shifts probabilities across the entire support, we select $q_{Sar}(x_i) = \exp(-x_i) - L_i(1)$, $x_i \in \mathbb{N}_0$, $i = 1,2$. Here, $L_i(1)$ is the value of the Laplace transform of the marginal distribution evaluated at $s=1$, that is
\[
L_i(s) = E\left(e^{-sX_i}\right)=\sum\limits_{x_i=0}^\infty \exp(-sx_i) P(x_i) \, ,
\]
with $P(\cdot)$ denoting the pmf of the $i-$th marginal distribution. Then, by plugging $q_{Sar}(x_i)$ into equation \ref{eq4}, a bivariate pmf is given by
\begin{equation*}
    P(X_1 = x_{1},X_2 = x_{2})=P_{1}(x_{1})P_{2}(x_{2})\left\{1+ \omega \left[ \exp(-x_1) - L_1(1) \right]
    \left[ \exp(-x_2) - L_2(1) \right] \right \}.
    \end{equation*}
In the following, we derive two bivariate pmfs according to this setup for Poisson and negative binomial marginals.

\subsubsection*{Example: Poisson margins}
Considering Poisson marginal distributions, we use the series expansion of the exponential function to derive the corresponding Laplace transform as $L_i(1)=\exp\big(-\lambda_i (1-\exp(-1))\big)$. The joint pmf is then given by 
 \begin{eqnarray*}
 P(X_1 = x_{1},X_2 = x_{2}) &=& \frac{{\lambda^{x_1}_1} \exp ( -\lambda_1 )}{x_1!}\frac{{\lambda^{x_2}_2} \exp (-\lambda_2 )}{x_2!} \times \nonumber \\ &&
\left\{ 1+ \omega \left[ (e^{-x_1} - e^{-\lambda_1 c})(e^{-x_2} -
e^{-\lambda_2 c})
\right] \right\}  
\end{eqnarray*} 
where $\omega$ is a dependence parameter, $\lambda_i,\ i = 1,2,$ are the means of the Poisson distributions and
$c=1-\exp(-1)$ is a constant. 
The bivariate Poisson distribution presented here has also been studied in \citet{lakshminarayana1999bivariate}.

\subsubsection*{Example: negative binomial margins}
Similar to the previous example, we can also derive the joint pmf for negative binomial margins as 
\begin{eqnarray*}
P(X_1 = x_1, X_2 = x_2) &=&
 \frac{\Gamma(x_1+\phi_1)}{\Gamma(\phi_1)x_1!}  \left( \frac{\phi_1}{\phi_1+\mu_1} \right)^{\phi_1}
 \left( \frac{\mu_1}{\phi_1+\mu_1} \right)^{x_1} \times \nonumber  \\
 &&
 \frac{\Gamma(x_2+\phi_2)}{\Gamma(\phi_2)x_2!}  \left( \frac{\phi_2}{\phi_2+\mu_2} \right)^{\phi_2}
 \left( \frac{\mu_2}{\phi_2+\mu_2} \right)^{x_2}
\times \nonumber \\
&&
\left\{ 1+ \omega \left[e^{-x_1} - L_1(1)\right]\left[e^{-x_2} - L_2(1)\right]
\right \}, 
\end{eqnarray*}
with $\mu_i$, denoting the mean of the $i$-th marginal distribution with the variances $\mu_i + \frac{\mu_i^2}{\phi_i}$, $i=1,2$. Here, $\phi_i$ constitutes the overdispersion parameter. The Laplace transform of the negative binomial distributions evaluated at $s = 1$, i.e. $L_i(1)$, for $i = 1,2$, is given by
\begin{equation*} L_i(1) = \left[ \frac{\phi_i}{\phi_i + \mu_i(1-e^{-1})}
\right]^{\phi_i}.
\end{equation*}
The distribution presented here in the second example is also examined by \citet{famoye2010bivariate}.

\begin{figure}
    \centering
    \includegraphics[width = 0.6\textwidth]{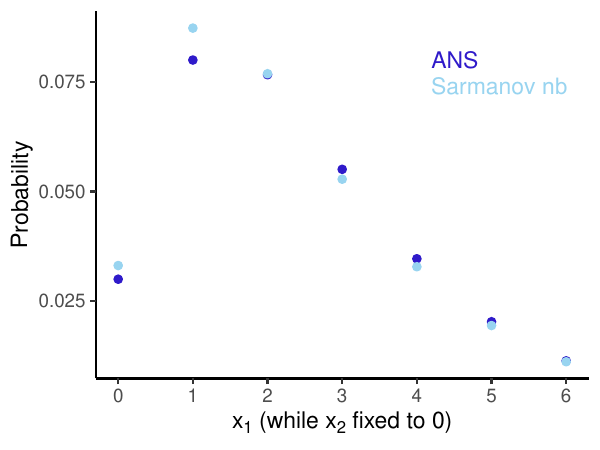}
   \caption{Probabilities (y-axis) under the ANS and the Sarmanov model with negative binomial marginals for $x_1$ (x-axis) while $x_2$ is set to zero. Parameters used for this figure were obtained from the fitted models in Section 3.2.}
   \label{fig:probs2}
\end{figure}

\subsubsection*{Example: Alternative Negative binomial Sarmanov model}
While the previous two examples have already been explored by \citet{lakshminarayana1999bivariate} and \citet{famoye2010bivariate}, respectively, we aim to generalise their distributions to end up with a more flexible model. To this end, by using
$q_{ANS}(x)~=~\big[\phi_i/(\phi_i + \mu_i)\big]^{x_i} - c_i$ (with $c_i$ defined below) for $i = 1,2$, we create a novel bivariate distribution. In fact, if $\phi_i/(\phi_i + \mu_i) = e^{-1}$, we obtain the bivariate distribution presented in the previous example. For the other cases, we create an alternative bivariate distribution with negative binomial margins based on the Sarmanov family without using the Laplace transform. We thus obtain the following pmf for the \textbf{A}lternative \textbf{N}egative binomial \textbf{S}armanov (ANS) distribution:
\begin{eqnarray*}
P(X_1 = x_1, X_2 = x_2) &=&
 \frac{\Gamma(x_1+\phi_1)}{\Gamma(\phi_1)x_1!}  \left( \frac{\phi_1}{\phi_1+\mu_1} \right)^{\phi_1}
 \left( \frac{\mu_1}{\phi_1+\mu_1} \right)^{x_1} \times \nonumber  \\
 &&
 \frac{\Gamma(x_2+\phi_2)}{\Gamma(\phi_2)x_2!}  \left( \frac{\phi_2}{\phi_2+\mu_2} \right)^{\phi_2}
 \left( \frac{\mu_2}{\phi_2+\mu_2} \right)^{x_2}
\times \nonumber \\
&& \left \{ 1+ \omega \left[\left(\frac{\phi_1}{\phi_1 + \mu_1}\right)^{x_1} - c_1\right]\left[\left(\frac{\phi_2}{\phi_2 + \mu_2}\right)^{x_2} - c_2\right]
\right \}
\end{eqnarray*}
where $\omega,\ \mu_i$ and $\phi_i,~i=1,2$ are parameters with the meaning as above and
\begin{equation*}  c_i = \left( \frac{\phi_i}{\phi_i + \mu_i} \right)^{\phi_i} \left[1-\left(1-\frac{\phi_i}{\phi_i + \mu_i}\right) \frac{\phi_i}{\phi_i + \mu_i} \right]^{-\phi_i}. \label{ccs2} \end{equation*}
In Appendix~\ref{sec:app_ANS}, we show that equation (\ref{cond}) still holds for this pmf. Correlation properties are reported in Appendix~\ref{sec:app_correlation}. 

Figure \ref{fig:probs2} illustrates the difference between the bivariate Sarmanov model using negative binomial marginals and the ANS model. Interpreted in terms of football scorelines, the ANS model shifts more weight from scoreless draws and close wins to clear wins compared to the Sarmanov model from the previous example.

\begin{table}
\tiny
\caption{\label{ratios} For each score, the joint frequency is divided by the product of the product of the marginal totals. The vertical numbers correspond to home goals whereas the horizontal numbers correspond to away goals. Standard errors obtained via bootstrapping are denoted in parenthesis.}
\begin{tabular}{rrrrrrrrrrrr}
  \hline
  & & \multicolumn{2}{c}{\textit{England}} & & & & & \multicolumn{2}{c}{\textit{Germany}} \\ 
 & \textbf{0} & \textbf{1} & \textbf{2} & \textbf{3} & \textbf{4} & & \textbf{0} & \textbf{1} & \textbf{2} & \textbf{3} & \textbf{4} \\ 
  \hline
\textbf{0}  & 0.74 (0.10) & 0.82 (0.10) & 1.16 (0.13) & 0.90 (0.20) & 1.82 (0.25)  &  \textbf{0}  & 0.46 (0.06) & 0.84 (0.07) & 1.04 (0.1) & 1.76 (0.15) & 2.19 (0.23) \\

  \textbf{1}  & 1.05 (0.11) & 0.95 (0.11) & 1.05 (0.14) & 1.01 (0.23) & 0.89 (0.24) & \textbf{1} & 0.83 (0.07) & 1.10 (0.08) & 1.17 (0.11) & 0.93 (0.14) & 0.85 (0.21)  \\ 

  \textbf{2}  & 0.87 (0.12) & 1.44 (0.13) & 0.78 (0.15) & 1.28 (0.29) & 0.34 (0.19) & \textbf{2} & 1.06 (0.09) & 1.08 (0.10) & 1.13 (0.14) & 0.91 (0.17) & 0.52 (0.20)  \\

  \textbf{3} & 1.19 (0.21) & 1.07 (0.20) & 0.86 (0.23) & 0.82 (0.37) & 0.94 (0.45) & \textbf{3} & 1.42 (0.13) & 1.00 (0.13) & 0.89 (0.17) & 0.69 (0.20) & 0.58 (0.27) \\   

  \textbf{4} & 1.67 (0.31) & 0.56 (0.22) & 1.07 (0.36) & 1.10 (0.58) & 0.42 (0.40) & \textbf{4} & 1.53 (0.16) & 1.17 (0.17) & 0.78 (0.2) & 0.10 (0.10) & 0.20 (0.20)  \\ 
  & &   \multicolumn{2}{c}{\textit{France}} & & & &  & \multicolumn{2}{c}{\textit{Spain}} \\
  & \textbf{0} & \textbf{1} & \textbf{2} & \textbf{3} & \textbf{4} & & \textbf{0} & \textbf{1} & \textbf{2} & \textbf{3} & \textbf{4} \\
  \textbf{0} & 0.44 (0.05) & 0.91 (0.07) & 0.93 (0.10) & 1.68 (0.15) & 1.94 (0.20)  & \textbf{0} & 0.66 (0.05) & 0.87 (0.06) & 1.09 (0.09) & 1.43 (0.12) & 2.06 (0.22) \\ 

  \textbf{1} & 0.73 (0.06) & 1.15 (0.09) & 1.22 (0.12) & 1.07 (0.15) & 1.19 (0.21)  & \textbf{1} & 0.85 (0.05) & 1.00 (0.05) & 1.12 (0.08) & 1.32 (0.11) & 1.06 (0.16) \\

  \textbf{2} & 1.05 (0.10) & 1.15 (0.12) & 1.12 (0.16) & 0.98 (0.21) & 0.55 (0.21)  & \textbf{2} & 0.98 (0.06) & 1.11 (0.06) & 1.06 (0.09) & 0.88 (0.12) & 0.61 (0.16) \\

  \textbf{3} & 1.48 (0.13) & 1.11 (0.15) & 0.97 (0.20) & 0.35 (0.17) & 0.29 (0.20) & \textbf{3} & 1.23 (0.08) & 1.09 (0.08) & 0.98 (0.11) & 0.43 (0.11) & 0.44 (0.17)  \\ 

  \textbf{4} & 1.54 (0.15) & 1.03 (0.17) & 1.04 (0.26) & 0.39 (0.22) & 0.00 (0.00)  & \textbf{4} & 1.59 (0.12) & 0.91 (0.12) & 0.58 (0.14) & 0.71 (0.20) & 0.29 (0.21) \\ 
 
   \hline
\end{tabular}
\end{table}

\section{Application}
\subsection{Data}
We fit the models introduced in the previous section to data from women's football. Specifically, we use data on the number of goals scored in each match in the seasons 2011/12-2018/19 and 2021/22 of the English FA Women's Super League, the German Frauen-Bundesliga, the French Division 1 Feminine and the Spanish Primera Iberdrola. We thus only consider matches from seasons that were not affected by COVID-19 restrictions.

\begin{figure}
    \centering
    \includegraphics[width=0.7\textwidth]{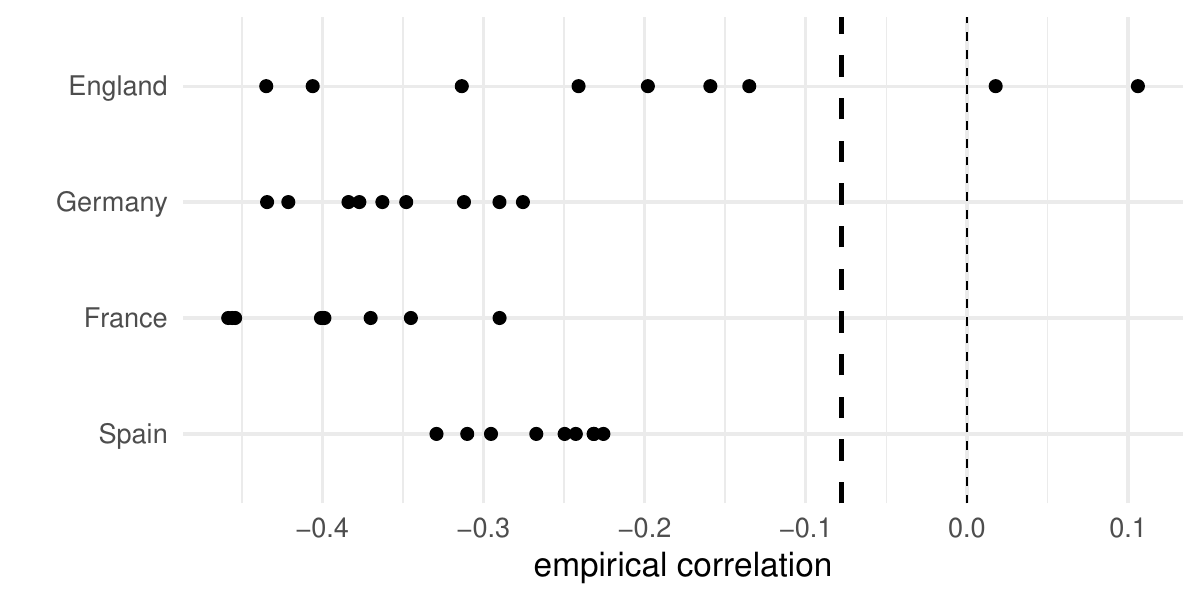}
    \caption{Empirical correlations found in the data of the goals scored by the home and away teams. Each dot represents one season. The black dashed line at $-0.08$ indicates the lower bound of the correlation implied
   under the classical Dixon and Coles model. $\lambda_1$ and $\lambda_2$ are set to 1.3 and 1.2, respectively, which are plausible values for football data. 
   Correlations slightly smaller than $-0.08$ can also be obtained under the Dixon and Coles model, but only for smaller means.}
    \label{fig:emp_corr}
\end{figure}

To investigate whether the number of home and away goals are independent, we consider their joint contingency table and calculate the ratio of the joint frequencies and the product of the marginal totals for each score. 
Table~\ref{ratios} displays these ratios for the most common results in all leagues, indicating that most ratios substantially differ from one --- thus, assuming independence of home and away goals does not seem appropriate. Chi-squared tests reject the null hypothesis of independence for each league except for the English FA Women's Super League (p-value: 0.067). 
While a dependence between home and away goals is in line with data from men's football (see, e.g., \citealp{dixon1997modelling, karlis2003analysis}), the ratios for women's football displayed in Table~\ref{ratios} indicate an \textit{under}representation of 0-0s in each league, whereas such scoreless draws are usually \textit{over}represented in men's football. Another common pattern in women's football is a substantial negative correlation between the two teams' number of goals which is an observation that bears similarities to those made by \cite{mchale2011modelling} in the context of international games. In particular, due to the underrepresentation of draws and overrepresentation of scores such as 3-0 and 4-0 in women's football, the correlations between home and away goals in our sample are -0.269 (England), -0.352 (Germany), -0.395 (France), and -0.263 (Spain). While for a few seasons, the correlations are positive, as shown in Figure \ref{fig:emp_corr}, they remain mostly negative for all leagues and seasons considered. More importantly, the relatively large amount of negative correlation could not be captured by the Dixon and Coles model --- for our data, the lower bound of the correlation is -0.05 (calculated based on the fitted models below). 
Figure \ref{images} in Appendix \ref{sec:app_correlation} 
shows the ranges of correlation implied by the Dixon and Coles model as well as the proposed extensions, indicating that the ANS model can capture a much wider range of correlation.

\begin{figure}[!h]
\centering
\includegraphics[width = 0.8\textwidth]{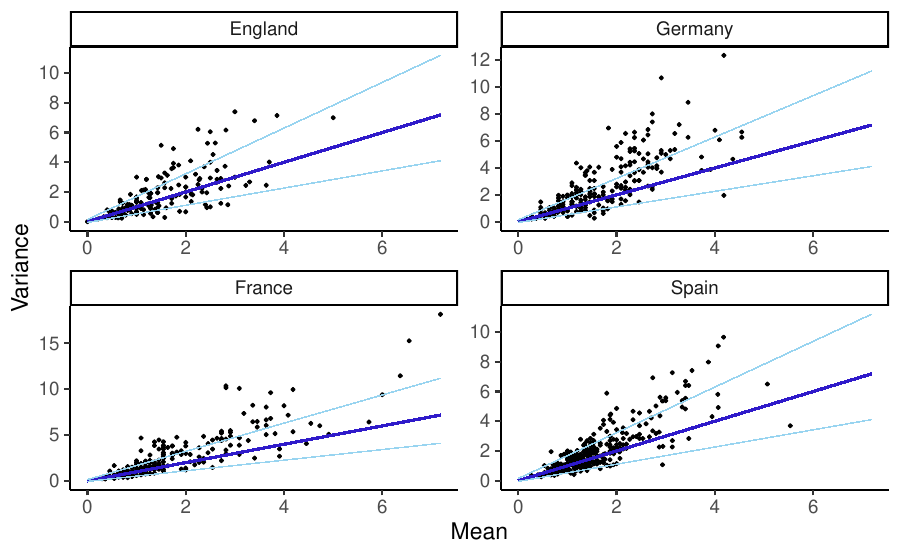}
\caption{\label{meanvar} The scatter plots display the mean and variance of the scores of the different teams in the seasons 2011/12 to 2018/19 and 2021/22, separated for home and away matches (i.e., two dots for each team). The blue line corresponds to the case where the mean equals the variance, and the light blue lines indicate corresponding 95\% confidence intervals obtained via Monte Carlo simulation.}
\label{mean_variance}
\end{figure}

Another pattern that is different in women's football compared to men's football concerns overdispersion. To illustrate this, Figure \ref{meanvar} shows the mean and variance of goals scored in home and away matches of the different teams. Here, one dot refers to a team's home/away performance, and the diagonal indicates mean-variance equivalence. 
For all four leagues, several data points lie above the diagonal, and for Germany, France, and Spain, several points also lie outside the 95\% confidence interval. Figure \ref{meanvar} thus indicates overdispersion in the data, potentially rendering the negative binomial distribution more suitable than the commonly used Poisson distribution for modelling the number of goals scored.
The remainder of this section first considers basic model formulations without including any covariates. To fully address the patterns observed in women's football scorelines, we employ the model formulations developed in Section~2. We further include team-specific attacking and defence parameters into the models and finally compare their fit and predictive performance.

\begin{table}[ht]
\centering
\caption{AIC for the models fitted to data from different leagues without team-specific effects. Values in bold indicate the models preferred by the AIC.}
  \setlength\extrarowheight{-3pt}
\begin{tabular}{lcccc}
  \hline
 & England & Germany & France & Spain \\ 
  \hline
  double Poisson & 4513.33 & 8873.91 & 9465.20 & 15432.49 \\ 
  double negative binomial & 4379.84 & 8337.95 & 8551.77 & 14953.09 \\
\hline
  Dixon and Coles Poisson & 4514.71 & 8874.65 & 9466.40 & 15434.23 \\ 
  Dixon and Coles Poisson with $\hat{q}$ & 4514.71 & 8874.65 & 9466.40 & 15434.23 \\ 
  Dixon and Coles Poisson with $\Tilde{q}$ & 4510.95 & 8874.01 & 9467.07 & 15432.01 \\ 
  Dixon and Coles negative binomial with $q_{nb}$ & 4380.81 & 8339.66 & 8553.77 & 14955.01 \\ 
  Dixon and Coles negative binomial with $\hat{q}_{nb}$ & 4380.81 & 8339.66 & 8553.77 & 14955.01 \\ 
  Dixon and Coles negative binomial with $\Tilde{q}_{nb}$ & 4377.38 & 8338.85 & 8552.51 & 14952.66 \\ 
  \hline
  Sarmanov Poisson & 4488.00 & 8780.07 & 9360.28 & 15352.42 \\ 
  Sarmanov negative binomial & 4347.13 & 8204.14 & 8372.73 & 14852.57 \\ 
  Alternative Negative binomial Sarmanov & \textbf{4332.66} & \textbf{8164.33} & \textbf{8343.46} & \textbf{14787.05} \\ 
   \hline
\end{tabular}
    \label{wodummies}
\end{table}

\subsection{Baseline model}
We first fit relatively simple models to the women's football data which do not include covariates and which we refer to as baseline models. To obtain parameter estimates, we numerically maximise the log-likelihood, which is carried out in R using the function \texttt{nlm()}. Table \ref{wodummies} displays the AIC values obtained for these models, indicating that the AIC prefers models with negative binomial margins over Poisson margins for each league. Specifically, among the models with negative binomial marginals, the ANS model is favoured by the AIC for all leagues. Compared to the other model formulations considered, the ANS model is more flexible and seems better to capture the specific dependence structure via its additional parameters. 

\subsection{Model including team dummies}
The model formulations presented next include team-specific effects. For all models, we use the same type of mean-parametrization as introduced in Section~2, i.e., the first parameter of a marginal distribution always represents the mean. To account for team-specific effects, we consider an attacking and a defence parameter for each team in each league. Additionally, we include a binary variable for home matches to account for the well-known home-field advantage (\citealp{karlis2003analysis,carmichael2005home,baker2006predicting}). In particular, for $n$ matches and two parameters $\theta_{1j}$ and $\theta_{2j}$, $j=1,\ldots,n$ representing the mean of the home and away team of the chosen distribution of the response variable, the linear predictors then have the following form
\begin{equation*}
\begin{split}
    \log(\theta_{1j}) & = \text{home} + \text{att}_{h_j} + \text{def}_{g_j}, \\
    \log(\theta_{2j}) & = \text{att}_{g_j} + \text{def}_{h_j},
\end{split}
\end{equation*}
where $\text{att}_k$ and $\text{def}_k$ denote the attacking and defence parameters of team $k$, with $k$ being a placeholder for either $h_j$ or $g_j$ to indicate the home and away team in match~$j$. To ensure identifiability, we use a sum-to-zero constraint for the defence parameters. 

\begin{table}[ht]
\centering
   \caption{AIC for the models fitted to data from different leagues including team-specific effects and a home dummy variable. Values in bold indicate the models preferred by AIC.}
     \setlength\extrarowheight{-3pt}
\begin{tabular}{lcccc}
  \hline
 & England & Germany & France & Spain \\ 
  \hline
double Poisson & 4016.23 & 7348.87 & 7110.91 & 13529.97 \\ 
double negative binomial & 4017.61 & 7334.03 & 7104.77 & 13518.99 \\ 
  \hline
  Dixon and Coles Poisson & 4018.07 & 7350.77 & 7112.88 & 13531.52 \\ 
  Dixon and Coles Poisson with $\hat{q}$ & 4018.10 & 7350.86 & 7112.68 & 13531.48 \\ 
  Dixon and Coles Poisson with $\Tilde{q}$ & \textbf{4014.14} & 7350.87 & 7112.86 & 13531.95 \\ 
  Dixon and Coles negative binomial with $q_{nb}$ & 4019.39 & 7335.94 & 7106.74 & 13520.57 \\ 
  Dixon and Coles negative binomial with $\hat{q}_{nb}$ & 4019.44 & 7336.00 & 7106.52 & 13520.51 \\ 
  Dixon and Coles negative binomial with $\Tilde{q}_{nb}$ & 4015.71 & 7336.02 & 7106.74 & 13520.99 \\ 
  \hline
  Sarmanov Poisson & 4016.31 & 7340.10 & 7112.18 & 13529.48 \\ 
  Sarmanov negative binomial & 4017.59 & 7324.81 & 7105.95 & 13518.47 \\ 
  Alternative Negative binomial Sarmanov & 4018.12 & \textbf{7321.30} & \textbf{7102.81} & \textbf{13515.88} \\ 
   \hline
\end{tabular}
    \label{AIC_incl}
\end{table}

Table \ref{AIC_incl} displays the AIC values obtained for the fitted models, including home and team-specific effects. While the AIC favours the ANS model for Germany, France, and Spain, the Dixon and Coles extended model shifting probabilities for scores up to 2 is preferred for England. We can explain this by the patterns in the data: the amount of overdispersion for the leagues in Germany, France, and Spain (cf.\ Figure \ref{mean_variance}) is larger than for England --- the AIC thus prefers the ANS model for these leagues. In contrast, since in the English FA Women's Super League only a few teams show overdispersion, the additional complexity of the ANS model is not required here.  

Our results suggest that the models developed in Section~2, especially the Alternative Negative binomial Sarmanov model, are more suitable for modelling the number of goals in women's football than classical models initially developed for men's football. 

\subsection{Model checking}
To check the adequacy of the preferred ANS model, we calculate the difference between the empirical proportions of scores and the probabilities under the different models. Table~\ref{tab:probs} displays the sum of the absolute differences for all models and leagues considered. For the English FA Women's Super League and the Spanish Primera Iberdrola, the Dixon and Coles model and an extension is preferred, respectively. However, the ANS model performs only slightly worse. 
At the same time, the ANS model shows the best model fit for the German Frauen-Bundesliga and the French Division 1 Feminine.
\begin{table}[h!]
\caption{Sums of the absolute differences between the probabilities (for the results 0-0, 1-0, 0-1, 1-1, \ldots, 11-11; other results' probabilities are approximately zero) under the different models and the corresponding empirical proportions. For clarity, we multiplied the differences by 100. Bold numbers indicate the models with the smallest absolute difference.}
\centering
  \setlength\extrarowheight{-3pt}
\begin{tabular}{lcccc}
  \hline
 & England & Germany & France & Spain \\ 
  \hline
  double Poisson & 19.80 & 13.98 & 13.72 &  9.58 \\ 
  double negative binomial & 19.21 & 13.76 & 14.42 & 11.26 \\ 
  \hline
  Dixon and Coles Poisson & 19.80 & 13.77 & 13.72 & \textbf{9.50} \\
  Dixon and Coles Poisson with $\hat{q}$ &  19.35 & 13.79 & 13.43 & 9.52 \\ 
  Dixon and Coles Poisson with $\Tilde{q}$ & 19.78 & 13.97 & 13.72 & 9.59 \\  
  Dixon and Coles negative binomial with $q_{nb}$ & \textbf{18.72} & 14.74 & 14.30 & 10.65 \\ 
  Dixon and Coles negative binomial with $\hat{q}_{nb}$ & 18.73 & 14.56 & 14.30 & 10.66 \\ 
  Dixon and Coles negative binomial with $\Tilde{q}_{nb}$ & 18.92 & 14.52 & 14.30 & 10.75 \\  
  \hline
  Sarmanov Poisson & 20.55 & 14.57 & 13.79 & 9.74 \\ 
  Sarmanov negative binomial & 19.56 & 13.96 & 13.76 & 10.48 \\ 
  Alternative Negative binomial Sarmanov & 19.19 & \textbf{13.19} & \textbf{13.28} & 9.61 \\ 
   \hline
\end{tabular}
\label{tab:probs}
\end{table}

\subsection*{Prediction}
\begin{figure}
    \centering
    \includegraphics[width=0.8\textwidth]{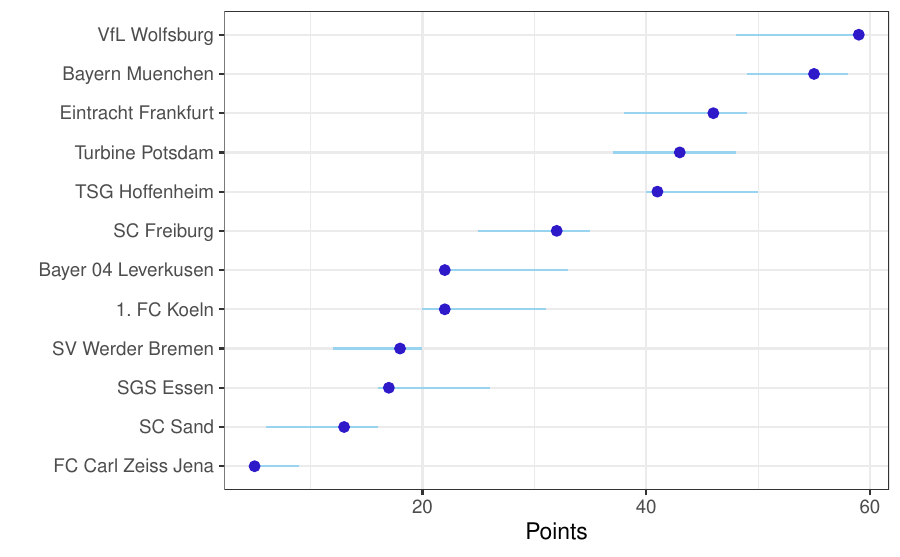}
   \caption{The plot displays the 95\% confidence intervals (light blue lines) of the simulated final tables of the German Frauen-Bundesliga under the ANS model. The dark blue points indicate the true final points of the teams.}
   \label{sims}
\end{figure}

To further demonstrate the usefulness of our approach in practice, we consider the predictive performance of the most promising model, the ANS model, for the German Frauen-Bundesliga. In particular, we fit the
ANS model to the first two-thirds of the season 2021/22, i.e.\ to the first 15 matchdays, and evaluate the predictive performance based on the remaining seven matchdays.
To this end, we predict the probabilities for all scores in each match under the fitted model. We then simulate all matches played in the last seven matchdays 1,000 times using a Monte Carlo simulation. In this way, we end up with 1,000 simulated final points of the teams. From these final points, we calculate the $2.5\%-$ and $97.5\%-$quantiles to obtain a 95\% prediction interval for each team. 
Figure \ref{sims} shows the observed standings at the end of the season, together with the predicted intervals as obtained under the ANS model. The predicted intervals include the observed final points for all teams, thus suggesting a promising predictive performance of the ANS model. 

\section{Discussion}
There is wide interest in predicting football matches by fans, media, and academics. To that end, we provide a modelling framework to model the number of goals scored in a football match, which is a flexible extension of the model developed by \citet{dixon1997modelling}. A vital strength of the proposed approach is that we can not only shift probabilities between scores in a very flexible way but also consider (discrete) marginal distributions other than Poisson. 

In our application, we tested the feasibility of our proposed models by analysing data from four different women's football leagues. In women's football, usually fewer 0-0s occur as expected under independence, which is different from men's football. To account for such characteristic in our modelling approach, we formulate several extensions of the Dixon and Coles model. The ANS model showed the most promising performance, both in terms of model fit and predictive power. Our application also demonstrated that the Dixon and Coles model is not able to capture all patterns in the data, especially the relatively large negative correlation found for women's football data.

For the prediction of scores in football, recent literature often uses several covariates, such as players' age and information about a team's recent performance (see, e.g., \citealp{groll2018dependency,whitaker2021bayesian}). For future research, our models could be extended to model one or multiple parameters via such covariates. In the presence of many covariates, regularisation approaches have proven helpful when predicting football results (\citealp{van2020generalised}). Moreover, as teams' performance may not be constant over time, a further model extension could include recent performance in a weighted manner by putting more weight on more recent observations --- \citet{dixon1997modelling} used a similar approach. As a team's form is usually not observable, the models presented here could also be extended by adding a latent state process. In particular, parameters such as the mean may depend on a team's underlying latent form. As the existing literature has already considered state-switching approaches in football \citep{otting2021copula},
such extensions could build upon the modelling framework developed in this contribution to flexibly model football scores, especially for women's football.

\bibliographystyle{apalike}

\newpage
\begin{appendices}

\section{Distributions with other exponents}\label{sec:app_other_exponents}
\numberwithin{equation}{section}
\setcounter{equation}{0}

As discussed in section 2.4.1, we can extend the model by \citet{dixon1997modelling} for other exponents than quadratic ones. To this end, assume a $q-$function of the form
\begin{equation}
    \label{newq2}
\hat{q}_{(s)}(x_i) = \left\{ \begin{array}{ll}
-\lambda_i^s & \text{if } x_i=0 \\
\lambda_i^{s-1} & \text{if }  x_i=1 \\
0 & \text{if } x_i=2,3,\ldots
\end{array}   \right.
\end{equation}
for $i= 1,2$ and again assume Poisson marginal distributions.
Then, based on equation \ref{eq4} from section 2.2, we have a bivariate distribution with pmf
\[
P(X_1 = x_1, X_2 = x_2) = \tau_{\lambda_1, \lambda_2}(x_1,x_2)\frac{{\lambda^{x_1}_1} \exp ( -\lambda_1 )}{x_1!}\frac{{\lambda^{x_2}_2} \exp (-\lambda_2 )}{x_2!},
\]
with
\[
\tau_{\lambda_1, \lambda_2}(x_1,x_2) = \left\{ \begin{array}{cc}
1 + \tilde{\omega} \lambda_1^s \lambda_2^s, \ & \mbox{if}~ x_1=x_2=0, \\
1 - \tilde{\omega} \lambda_1^s \lambda_2^{s-1}  \ & \mbox{if}~ x_1=0, x_2=1, \\
1 - \tilde{\omega} \lambda_1^{s-1} \lambda_2^s    \ & \mbox{if}~ x_1=1, x_2=0, \\
1 + \tilde{\omega} \lambda_1^{s-1} \lambda_2^{s-1} & \mbox{if}~ x_1=x_2=1, \\
1 & \mbox{otherwise}.
\end{array}
\right.
\]
\noindent Obviously for $s=1$ we get the Dixon and Coles model.
In order for the above to be a proper pmf it must hold that 
$\tau_{\lambda_1, \lambda_2}(x_1,x_2) \ge 0$ and then we can derive the
restrictions that
\[
\max \left( -\frac{1}{\lambda_1^s \lambda_2^s}, 
-\frac{1}{\lambda_1^{s-1} \lambda_2^{s-1}}  \right)
\le \tilde{\omega} \le \min \left( -\frac{1}{\lambda_1^s \lambda_2^{s-1}},
-\frac{1}{\lambda_1^{s-1}\lambda_2^s }  \right) 
\]
Note that $\tilde{\omega}$ is a parameter with different interpretation of each $s$.

\section{Proof: Alternative Negative binomial Sarmanov model}\label{sec:app_ANS}
Here, we demonstrate that the ANS distribution from section 2.5 indeed fulfils equation \ref{cond} from section 2.2. For this, we have to proof that the expectation of the corresponding $q-$function, $q_{ANS}$, equals zero, i.e.:

Theorem 1:

Let $q_{ANS}(x_i) = \left(\frac{\phi_i}{\phi_i + \mu_i}\right)^{x_i} - c_i$ with 
\begin{equation*}  c_i = \left( \frac{\phi_i}{\phi_i + \mu_i} \right)^{\phi_i} \left[1-\left(1-\frac{\phi_i}{\phi_i + \mu_i}\right) \frac{\phi_i}{\phi_i + \mu_i} \right]^{-\phi_i}. \label{ccs} \end{equation*} 
Then $\int\limits_{x_i=-\infty}^{\infty} q_{i}(x_{i})f_{i}(x_{i})=0$, for $i = 1,2$.

\textit{Proof:}
If E$\left[\left(\frac{\phi_i}{\phi_i + \mu_i}\right)^{x_i}\right] = c_i$, then

\begin{align*}
    \int\limits_{x_i=-\infty}^{\infty} q_{i}(x_{i})f_{i}(x_{i}) &= 
    \sum_{x_i = 0}^{\infty}\left[\left(\frac{\phi_i}{\phi_i + \mu_i}\right)^{x_i}-c_i\right] P(x_i) \\
    &= \sum_{x_i = 0}^{\infty}\left[\left(\frac{\phi_i}{\phi_i + \mu_i}\right)^{x_i}\right] P(x_i) - c_i \sum_{x_i = 0}^{\infty} P(x_i) \\
    &= E\left[\left(\frac{\phi_i}{\phi_i + \mu_i}\right)^{x_i}\right] - c_i \\
    &= E\left[\left(\frac{\phi_i}{\phi_i + \mu_i}\right)^{x_i}\right] - E\left[\left(\frac{\phi_i}{\phi_i + \mu_i}\right)^{x_i}\right] = 0
\end{align*} for $i = 1,2.$
In the first equation, we plug in the definition of $q(x_i)$ and $f(x_i)$. As $c_i$ is a constant we can split the sum and drag $c_i$ out of the sum. As $P(\cdot)$ is a probability function the sum over all possible values is $1$.

For $i =1,2,$, it remains to show that E$\left[\left(\frac{\phi_i}{\phi_i + \mu_i}\right)^{x_i}\right] = c_i$:
\begin{align*}
    E\left[\left(\frac{\phi_i}{\phi_i + \mu_i}\right)^{x_i}\right] &= \sum_{x_i = 0}^{\infty}\left(\frac{\phi_i}{\phi_i + \mu_i}\right)^{x_i} P(x_i) \\
    &= \sum_{x_i = 0}^{\infty}\left(\frac{\phi_i}{\phi_i + \mu_i}\right)^{x_i} \binom{x_i + \phi_i - 1}{x_i}\left(\frac{\phi_i}{\phi_i + \mu_i}\right)^{\phi_i} \left (1-\frac{\phi_i}{\phi_i + \mu_i}\right)^{x_i} \\
    &= \sum_{x_i = 0}^{\infty} (-1)^{x_i} \binom{-\phi_i}{x_i} \left(\frac{\phi_i}{\phi_i + \mu_i}\right)^{\phi_i} \left(\frac{\phi_i}{\phi_i + \mu_i}\right)^{x_i} \left (1-\frac{\phi_i}{\phi_i + \mu_i}\right)^{x_i} \\
    &= \left(\frac{\phi_i}{\phi_i + \mu_i}\right)^{\phi_i} \sum_{x_i = 0}^{\infty} \binom{-\phi_i}{x_i} \left[-\left(\frac{\phi_i}{\phi_i + \mu_i}\right)\left (1-\frac{\phi_i}{\phi_i + \mu_i}\right)\right]^{x_i} \\
    &= \left(\frac{\phi_i}{\phi_i + \mu_i}\right)^{\phi_i} \left \{1+\left[-\left(\frac{\phi_i}{\phi_i + \mu_i}\right)\left (1-\frac{\phi_i}{\phi_i + \mu_i}\right)\right]\right \}^{-\phi_i} \\
    &= \left( \frac{\phi_i}{\phi_i + \mu_i} \right)^{\phi_i} \left[1-\left(1-\frac{\phi_i}{\phi_i + \mu_i}\right) \frac{\phi_i}{\phi_i + \mu_i} \right]^{-\phi_i} = c_i
\end{align*} \hfill $\Box$

We first plug in the definition of the expectation. Then, we use the definition of the pmf of the negative binomial distribution. After rearranging terms and dragging out the constant $\left( \frac{\phi_i}{\phi_i + \mu_i} \right)^{\phi_i}$, we can use the binomial series as the inner part of the square brackets is always smaller than $1$ in absolute value. Thus, we end up with the definition of $c_i$.

\section{Correlation for the ANS model}\label{sec:app_correlation}
While the calculation of the correlation for most of the models considered in this paper is straightforward, the calculation of the correlation for the ANS model from section 2.5 is a little bit more challenging. Thus, we outline it here:
Consider the negative binomial distribution with pmf
\[
P(X=k) = \frac{\Gamma(k+r)}{k! \Gamma(r)} (1-p)^r p^k,~~k=0,1,2,\ldots,~~r>0,~p \in (0,1).
\]
We can see that for this distribution it holds that
\[
E(X t^X ) = \left( \frac{1-p}{1-pt} \right)^r \left( \frac{ptr}{1-pt} \right).
\]
To see that, it suffices to note that it holds from the pmf of the negative binomial that
\[
\sum\limits_{k=0}^\infty \frac{\Gamma(k+r)}{k!}  p^k = \frac{\Gamma(r)}{(1-p)^r}.
\]
Taking derivative with respect to $p$ we derive that
\[
\sum\limits_{k=0}^\infty  k p ^{k-1} \frac{\Gamma(k+r)}{k!}  = \frac{r \Gamma(r)}{(1-p)^{r+1}}.
\]
Thus, for the expectation of interest we have that
\begin{eqnarray*}
E(Xt^X) &=& \sum\limits_{k=0}^\infty  k t^k \frac{\Gamma(k+r)}{k! \Gamma(r)} (1-p)^r p^k  \\
 &=&  \frac{(1-p)^r pt}{\Gamma(r)} \sum\limits_{k=0}^\infty  k \frac{\Gamma(k+r)}{k!}  (pt)^{k-1} \\
 &=&  \left( \frac{1-p}{1-pt} \right)^r \left( \frac{ptr}{1-pt} \right).
\end{eqnarray*}
Consider now the Alternative Negative binomial Sarmanov model with
\[
q_{ANS}(x_i) = \left(\frac{\phi_i}{\phi_i + \mu_i}\right)^{x_i} - c_i
\]
where 
\begin{equation*}  c_i = \left( \frac{\phi_i}{\phi_i + \mu_i} \right)^{\phi_i} \left[1-\left(1-\frac{\phi_i}{\phi_i 
+ \mu_i}\right) \frac{\phi_i}{\phi_i + \mu_i} \right]^{-\phi_i}. \end{equation*}
Hence, for $i = 1,2,$
\[
E\left[X_iq_{ANS}(X_i)\right]= E\left[ X_i \left( \frac{\phi_i}{\phi_i+\mu_i} \right)^{X_i}\right] - c_i E(X_i) 
\]
and after using the expectation representation from above and tedious algebraic manipulations we end up with
\[
E[X_iq_{ANS}(X_i)]= \left[ \frac{\phi_i(\mu_i+\phi_i)}{(\mu_i+\phi_i)^2 - \mu_i \phi_i)} \right]^{\phi_i} 
\left[ \frac{\mu_i \phi_i^2}{(\mu_i+\phi_i)^2 - \mu_i \phi_i)} -\mu_i\right].
\]

\renewcommand\thefigure{\thesection\arabic{figure}}
\setcounter{figure}{0}

\begin{figure}
    \centering
    \includegraphics[width=\textwidth]{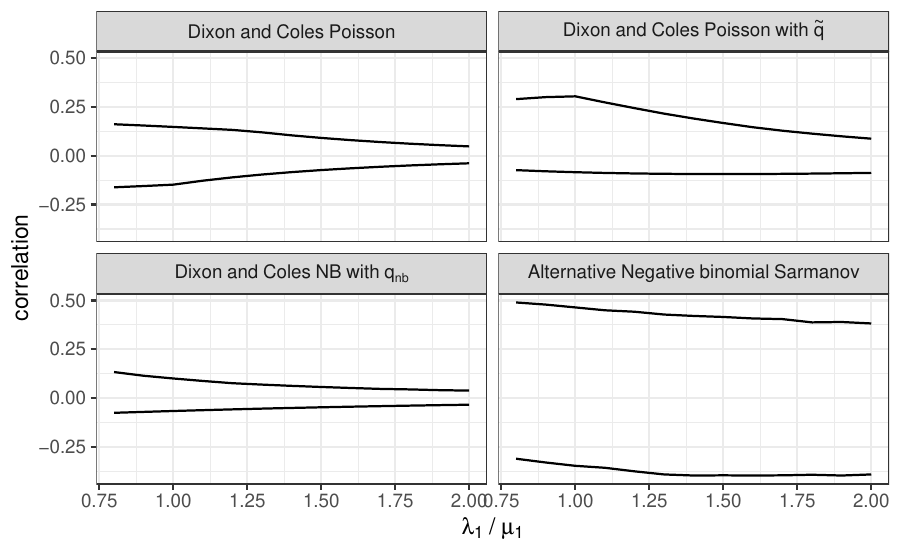}
    \caption{\label{images}Range of correlation for different models. What we depict here is the minimum and maximum value as a function of, in the top plots, $\lambda_1$ for all values of $\lambda_2$ and in the bottom plots $\mu_1$ for all values of $\mu_2$, $\phi_1$ and $\phi_2$. The ANS model widens the range of possible correlations and can thus adequately capture the empirical correlations shown in Figure~5.}
    
\end{figure}

\end{appendices}
\end{spacing}
\end{document}